\documentclass[aps,prab,twocolumn,showpacs,superscriptaddress,longbibliography]{revtex4-2}

\usepackage{graphicx}
\usepackage{amsmath}
\usepackage{amssymb}
\usepackage{bm}
\usepackage{booktabs}
\usepackage{siunitx}
\usepackage[colorlinks=true,linkcolor=blue,citecolor=blue,urlcolor=blue]{hyperref}

\newcommand{\A}{\mathcal{A}}
\newcommand{\B}{\mathcal{B}}
\newcommand{\dd}{\mathrm{d}}

\begin{document}

\title{Distinguishing the Axion-Wind and Oscillating-EDM Couplings\\
       in Storage-Ring Searches for Axion Dark Matter}

\author{V.\,Hejny}
\affiliation{Institut f\"ur Kernphysik, Forschungszentrum J\"ulich, 52425 J\"ulich, Germany}
\author{J.\,Pretz}
\affiliation{III. Physikalisches Institut B, RWTH Aachen University, 52056 Aachen, Germany}
\affiliation{Institut f\"ur Kernphysik, Forschungszentrum J\"ulich, 52425 J\"ulich, Germany}	


\begin{abstract}
Resonant spin-precession experiments in storage rings are sensitive to two distinct axion and axion-like particle dark-matter couplings: an oscillating electric dipole moment (EDM) primarily driven by the axion-gluon interaction, and the axion-wind (derivative) coupling acting as an effective pseudomagnetic field on the spin.  
Both effects produce a vertical polarization buildup at the same resonance frequency and are therefore degenerate in this standard observable.  
It is shown that the two couplings enter the vertical build-up and the spin-precession phase walk as the difference and sum of their resonance strengths, respectively.  
Measuring both observables simultaneously at fixed energy in a multi-bunch configuration allows the individual coupling strengths to be extracted
including their relative sign. Only the overall sign, equivalent to a shift of the axion phase by $\pi$, remains undefined.
As a complementary approach, the ratio of electric to magnetic bending fields in a combined storage ring can be varied while keeping the resonance frequency fixed.  This leaves the axion-wind contribution unchanged while modifying the EDM contribution in a calculable way, enabling a clean separation of the two effects.  
Both approaches can be used in future storage-ring facilities dedicated to EDM measurements.
\end{abstract}

\maketitle

\section{Introduction}
\label{sec:intro}

Axions arise naturally in extensions of the Standard Model addressing the strong-$CP$ problem~\cite{Peccei1977,Peccei1977b,Weinberg1978,Wilczek1978}.  
Axion-like particles (ALPs) are a broader class of light pseudoscalar bosons with similar phenomenology. 
Both are compelling dark-matter candidates and are referred to collectively as axions throughout this paper.
Their oscillating background field couples to ordinary matter through two qualitatively distinct spin interactions, both of which produce resonant signatures in spin-precession experiments.
The first is an oscillating contribution $d_{\rm ac}$ to the electric dipole moment (EDM),
\begin{equation}
  d(t) = d_{\rm dc} + d_{\rm ac}\cos(\omega_a t + \phi_a)\,,
  \label{eq:edm_osc}
\end{equation}
induced primarily by the axion-gluon interaction and proportional to the axion field 
\begin{equation}
  a(t) = a_0\cos(\omega_a t + \phi_a)\,,
  \label{eq:axion_field}
\end{equation}
such that $d_{\rm ac} \propto a_0$.  
The second is the axion-wind (derivative) coupling, proportional to $\partial_\mu a$, which acts on nuclear spins as an effective oscillating pseudomagnetic field aligned with the direction of dark-matter
flow~\cite{Graham2013, Stadnik2014}.

The two couplings have been targeted by dedicated laboratory programmes.  
The Cosmic Axion Spin Precession Experiment (CASPEr) utilizes nuclear magnetic resonance (NMR) techniques to search for axion dark matter by detecting its interaction with nuclear spins~\cite{Budker2014, JacksonKimball2017}. 
The EDM coupling necessitates the presence of an external electric field to induce a detectable torque on the spins, which is investigated by the CASPEr Electric branch employing ferroelectric crystals to provide strong internal electric fields. 
In contrast to that, the axion wind coupling acting as an oscillating pseudo-magnetic field interacts directly with nuclear magnetic moments without requiring an external electric field. The CASPEr Wind branch explores this coupling primarily using hyperpolarized liquid $^{129}$Xe across various magnetic field strengths, as well as Zero-to-Ultralow Field (ZULF) NMR methods that utilize specialized hyperpolarization techniques~\cite{Garcon2019,Wu2019}.

The nEDM experiment at PSI~\cite{Abel2017} exploited the fact, that the oscillating-EDM signal scales with the applied electric field $E$ while the axion-wind signal does not. 
The collaboration analyzed the ratio of ultracold-neutron and $^{199}$Hg spin-precession frequencies under parallel and antiparallel electric and magnetic field configurations allowing the two contributions to be separated.

In laboratory experiments the axion-wind signal is usually suppressed by the small non-relativistic velocity $v_\odot \approx 10^{-3}c$ of the Solar System with respect to the presumed galactic dark-matter halo.  Because $\vec v_\odot$ points in a fixed direction in the galactic frame,
the signal exhibits a characteristic sidereal modulation as the orientation of the laboratory quantization axis rotates with the Earth, producing sidebands at $\omega_a \pm \omega_{\rm sid}$ that are absent from the oscillating-EDM signal~\cite{Abel2017,Wu2019}.

Storage-ring experiments searching for ALP dark matter exploit the resonant coupling of the beam polarization to an oscillating dark-matter field~\cite{Karanth2023,Graham2021,Silenko2022,Nikolaev2022}.
The first search was performed at the Cooler Synchrotron (COSY) by the JEDI Collaboration using an in-plane polarized deuteron beam, scanning the spin-precession frequency through a 1.5~kHz window while monitoring the vertical polarization buildup as the signature of a resonant axion coupling~\cite{Karanth2023}.

For relativistic beams with $\beta\approx 1$ the axion-wind contribution is enhanced by a factor of order $\beta/\beta_\odot\sim 10^3$ relative to laboratory-frame experiments.  
However, unlike in laboratory experiments, the two handles available there --- electric-field reversal and sidereal modulation --- are not directly accessible in storage rings. The direction of $\vec{\beta}$ rotates continuously around the ring, averaging out any sidereal modulation within a single revolution.  
In the applied scanning method, both couplings contribute to the vertical polarization buildup at the same resonance frequency and with the same time dependence, so that only a single combined resonance strength can be extracted ~\cite{Karanth2023}.  Separate bounds on each coupling then require assuming that the other vanishes.

The present paper identifies two strategies specific to storage rings that break this degeneracy and derives the quantitative relations required to extract the individual coupling strengths.  
Section~\ref{sec:dynamics} establishes the equations of motion, Sec.~\ref{sec:eom} derives the build-up and phase-walk observables, and Secs.~\ref{sec:disentangle} and~\ref{sec:EB} present the two disentanglement strategies.

\section{Spin dynamics in the presence of EDM and axion-wind couplings}
\label{sec:dynamics}

\begin{figure}
  \includegraphics[width=\columnwidth]{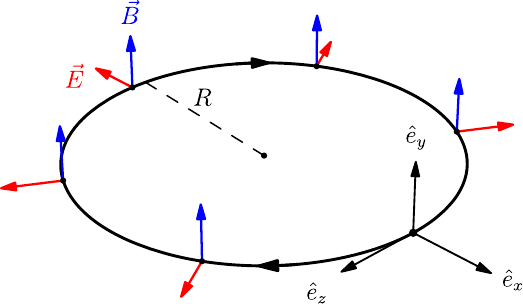}
  \caption{Coordinate system used throughout the paper: $\hat{e}_x$ points radially
    outward, $\hat{e}_y$ vertically upward, and $\hat{e}_z$ along the beam
    direction. The magnetic field $\vec{B}$ is vertical and the
    electric field $\vec{E}$ is radial. $R$ is the ring radius.}
  \label{fig:coordsys}
\end{figure}

The spin evolution in a storage ring is governed by the extended Thomas--BMT equation~\cite{Silenko2022,Nikolaev2022}
\begin{equation}
  \frac{\dd\vec{S}}{\dd t} =
    \bigl(\vec{\Omega}_{\rm MDM} - \vec{\Omega}_{\rm rev}
        + \vec{\Omega}_{\rm EDM} + \vec{\Omega}_{\rm wind}\bigr)
    \times \vec{S}\,,
  \label{eq:BMT}
\end{equation}
with the angular velocities
\begin{align}
  \vec{\Omega}_{\rm MDM} &= -\frac{q}{m}
    \left[
      \left(G+\frac{1}{\gamma}\right)B\,\hat{e}_y
      - \left(G+\frac{1}{\gamma+1}\right)\frac{\beta E}{c}\,\hat{e}_y
    \right],
  \label{eq:OMDM}\\[4pt]
  \vec{\Omega}_{\rm rev} &= \frac{q}{m\gamma}
    \left[\frac{\beta E}{\beta^2 c}\,\hat{e}_y - B\,\hat{e}_y\right],
  \label{eq:Orev}\\[4pt]
  \vec{\Omega}_{\rm EDM} &= \frac{d(t)}{S\hbar}
    \bigl(E - c\beta B\bigr)\hat{e}_x\,,
  \label{eq:Oedm}\\[4pt]
  \vec{\Omega}_{\rm wind} &= \frac{C_N}{2f_a S}
    \bigl(\hbar\partial_0 a(t)\bigr)\beta\,\hat{e}_z\,.
  \label{eq:Owind}
\end{align}
Here $\hat{e}_x$, $\hat{e}_y$, $\hat{e}_z$ point radially outward, vertically upward, and along the beam direction, respectively. $G$ is the anomalous magnetic moment, $\vec{B} = B\hat{e}_y$ and $\vec{E} = E\hat{e}_x$
are the vertical magnetic and radial electric fields (see Fig.~\ref{fig:coordsys}). $C_N/f_a$ is the
axion--nucleon coupling constant, and $\beta$, $\gamma$ are the usual relativistic
factors.  Terms involving $\vec\beta\cdot \vec{B}$ and $\vec\beta\cdot \vec{E}$ are omitted.  The minus sign in Eq.\eqref{eq:Oedm} arises from $\vec\beta\times\vec{B} = -\beta B\hat{e}_x$. The
permanent EDM component $d_{\rm dc}$ in Eq.~\eqref{eq:edm_osc} does not affect the following discussion and is omitted throughout.

The precession and revolution angular velocities combine to
\begin{eqnarray}
  \vec{\Omega}_{\rm prec} &\equiv& \vec{\Omega}_{\rm MDM} - \vec{\Omega}_{\rm rev} \\
  &=& -\frac{q}{m}\left[GB - \left(G - \frac{1}{\gamma^2-1}\right)
    \frac{\beta E}{c}\right]\hat{e}_y\,.
  \label{eq:Oprec}
\end{eqnarray}
Using the axion field $a(t)$, Eq.~\eqref{eq:axion_field}, the two ALP-induced contributions to the spin-precession vector become
\begin{align}
  \vec{\Omega}_{\rm EDM}  &= \frac{d_{\rm ac}}{S\hbar}(E-c\beta B)
    \cos(\omega_a t+\phi_a)\,\hat{e}_x\,,
  \label{eq:OedmALP}\\[4pt]
  \vec{\Omega}_{\rm wind} &= -\frac{C_N}{2f_a S}\,a_0\omega_a\beta\,
    \sin(\omega_a t+\phi_a)\,\hat{e}_z\,.
  \label{eq:OwindALP}
\end{align}
Notably, the rotation axes of the EDM and wind contributions are orthogonal ($\hat{e}_x$ vs. $\hat{e}_z$) and differ by a phase shift of $\pi/2$ (cosine vs. sine). 
This follows because the wind torque is proportional to $\partial_t a(t)$ and the EDM torque to $a(t)$
itself, so the two are $\pi/2$ out of phase.
Both torques nevertheless project onto the vertical direction at resonance and produce indistinguishable polarization buildups in the standard observable.

The resonance condition requires the axion oscillation frequency to match  the spin-precession frequency:
\begin{equation}
  \vec{\Omega}_{\rm prec} = \omega_a\,\hat{e}_y\,,
  \label{eq:resonance_vec}
\end{equation}
where $\omega_a$ is a signed frequency: its magnitude relates to the axion mass via $\hbar|\omega_a| = m_a c^2$, while its sign is fixed by the direction of the spin precession. This ensures consistency for particles with either sign of $G$.
Furthermore, from the EDM and axion-wind rotation vectors, Eqs.~\eqref{eq:OedmALP} and~\eqref{eq:OwindALP}, one can verify
that the simultaneous substitution $\omega_a \to -\omega_a$, $\phi_a \to -\phi_a$ leaves both torques unchanged.
Since the axion phase $\phi_a$ is unknown and is ideally probed using multi-bunch operation, the sign of $\omega_a$ is not directly observable from the EDM and wind signals alone. However, the sign of $\omega_a$ is physically bound through the resonance condition: a wrong sign would make the spin-precession frequency counter-rotate with respect to the axion-induced torques, turning the resonant enhancement into a destructive interference. 

\section{Build-up and phase walk: equations of motion}
\label{sec:eom}

\begin{figure}
  \includegraphics[width=\columnwidth]{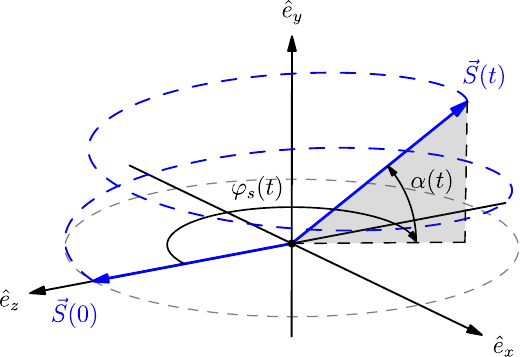}
  \caption{Definition of the out-of-plane angle $\alpha(t)$ and the in-plane spin-precession phase $\varphi_s(t)$. The relative phase $\varphi(t)$ is measured relative to the axion phase and is  $\varphi_s(t) = \omega_a t +\varphi(t)$ (see Eq.~\ref{eq:ansatz}). }
  \label{fig:alpha_phi}
\end{figure}

The dynamics is conveniently expressed in units of spin rotations per particle revolution. 
The axion-induced angular velocities can be written as
\begin{align}
  \vec{\Omega}_{\rm EDM}  &= 4\pi f_{\rm rev} \epsilon_{\rm ac}
    \cos(\omega_a t +\phi_a)\,\hat{e}_x\,,\\[4pt]
  \vec{\Omega}_{\rm wind} &= 4\pi f_{\rm rev} \epsilon_{\rm wind}
    \sin(\omega_a t+\phi_a)\,\hat{e}_z\,.
\end{align}
with the resonance tunes $\epsilon_{\rm ac}$ and $\epsilon_{\rm wind}$:
\begin{align}
  \epsilon_{\rm ac}   &= \frac{d_{\rm ac}}{S\hbar}
    \frac{(E-c\beta B)}{4\pi f_{\rm rev}}\,,
  \label{eq:eac}\\[4pt]
  \epsilon_{\rm wind} &= -\frac{C_N}{2f_a S}\,a_0\omega_a\beta
    \frac{1}{4\pi f_{\rm rev}}\,.
  \label{eq:ewind}
\end{align}
The factor $4\pi$ — rather than $2\pi$ — accounts for the factor of $1/2$ that
arises from averaging the oscillating torque over a full precession period~\cite{Eversmann2016}.
The spin motion can then be described using the ansatz
\begin{equation}
  \vec{S}(t) = S
  \begin{pmatrix}
    -\cos\alpha(t)\sin(\omega_a t+\varphi(t))\\
     \sin\alpha(t)\\
     \cos\alpha(t)\cos(\omega_a t+\varphi(t))
  \end{pmatrix},
  \label{eq:ansatz}
\end{equation}
where $\alpha(t)$ is the out-of-plane angle and $\varphi(t)$ is the
relative phase between the spin-precession and the oscillation of the axion field (see Fig.~\ref{fig:alpha_phi}). They evolve on a timescale 
much longer than one revolution.
Inserting Eq.~\eqref{eq:ansatz} into Eq.~\eqref{eq:BMT} and averaging over many
complete precession periods, one obtains
\begin{align}
  \frac{\dd\alpha}{\dd t}   &= 2\pi f_{\rm rev} (\epsilon_{\rm wind}-\epsilon_{\rm ac})
    \cos(\varphi-\phi_a)\,,
  \label{eq:dalpha}\\[6pt]
  \frac{\dd\varphi}{\dd t}  &= -2\pi f_{\rm rev} \tan\alpha\,
    (\epsilon_{\rm wind}+\epsilon_{\rm ac})\sin(\varphi-\phi_a)\,.
  \label{eq:dphi}
\end{align}
These equations are structurally analogous to those derived for resonant spin-flipper
devices~\cite{Eversmann2016} and exhibit a clear separation: the vertical
build-up $\dd\alpha/\dd t$ is proportional to the difference
$(\epsilon_{\rm wind}-\epsilon_{\rm ac})$, while the spin-precession phase walk $\dd\varphi/\dd t$
is proportional to the sum $(\epsilon_{\rm wind}+\epsilon_{\rm ac})$.

It should be noted that an rf Wien filter or rf solenoid operated as additional spin rotator at the spin-precession resonance produces additional torques along $\hat{e}_x$ and $\hat{e}_z$, respectively, \emph{i.e.} around the same axes and with the same $\pi/2$ phase relationship as the EDM and axion-wind torques. Each device therefore only adds to $\dd\alpha/\dd t$ and $\dd\varphi/\dd t$ in exactly the same combination of $\epsilon_{\rm wind}$ and $\epsilon_{\rm ac}$ and cannot be used to disentangle the oscillating-EDM and axion-wind contributions.

\section{Disentanglement via simultaneous build-up and phase-walk measurement}
\label{sec:disentangle}

Equations~\eqref{eq:dalpha} and~\eqref{eq:dphi} show that the build-up and the phase
walk are sensitive to complementary combinations of the two resonance tunes.  
Both observables are accessible with techniques demonstrated at COSY. The in-plane
precession phase and the spin tune are reconstructed from time-stamped polarimeter events~\cite{Eversmann2016}. This technique was used in feedback systems to stabilize the precession phase relative to an external reference~\cite{JEDI:2017bnp} or the external reference itself~\cite{JEDI:2025eec}. It was also used to analyze the phase walk for rf-solenoid driven spin oscillations~\cite{JEDI:2018txs}.

If build-up and phase walk are measured simultaneously, the signed combinations
\begin{equation}
  \A \equiv \epsilon_{\rm wind}-\epsilon_{\rm ac}\,,\qquad
  \B \equiv \epsilon_{\rm wind}+\epsilon_{\rm ac}
  \label{eq:AB}
\end{equation}
can be extracted by fitting the $(\varphi-\phi_a)$ dependence in a multi-bunch
configuration where different bunches carry known relative phase
offsets~\cite{Karanth2023}.
$\A$ enters Eq.~\eqref{eq:dalpha} with $\cos(\varphi-\phi_a)$ and $\B$ enters
Eq.~\eqref{eq:dphi} with $\sin(\varphi-\phi_a)$. A simultaneous fit of both
equations with a common axion phase $\phi_a$ therefore fixes the relative sign of
$\A$ and $\B$. The only remaining symmetry of the fit is
$(\A,\B,\phi_a)\to(-\A,-\B,\phi_a+\pi)$.
The individual resonance tunes
\begin{equation}
  \epsilon_{\rm wind} = \tfrac{1}{2}(\B+\A)\,,\qquad
  \epsilon_{\rm ac}   = \tfrac{1}{2}(\B-\A)
  \label{eq:eps_from_AB}
\end{equation}
are thus determined in magnitude and relative sign. Their overall sign remains
undefined, since it corresponds to a shift of the unknown axion phase by $\pi$.

This strategy requires a fixed-energy run at resonance, so that both
$\alpha(t)$ and $\varphi(t)$ can be tracked over many turns.  The scan method
of Ref.~\cite{Karanth2023} --- where the resonance is crossed in a time short compared
to the axion coherence time $\tau_a$
--- produces an essentially
instantaneous polarization jump whose amplitude is determined by the total resonance
strength alone, with no independent access to the sum and
difference.  A dedicated fixed-energy session is therefore required.  

Its feasibility rests on two conditions: the run duration must be short
compared to both $\tau_a$, so that the axion phase remains coherent, and
the in-plane polarization lifetime, so that $\alpha(n)$ and $\varphi(n)$
can be observed with adequate statistics.
Using $\tau_a = h/(m_a v^2)$ (see Eq.~(5) in Ref.~\cite{Karanth2023}) with $v \approx10^{-3}c$
(the velocity of the solar system relative to the assumed galactic axion field),
a minimum run duration of \SI{10}{s} restricts the accessible mass range to
$m_a \leq \SI{0.4}{neV/c^2}$.
The corresponding coherence length $l_a = h/(m_a v)$ exceeds any
realistic ring diameter, so the spatial extent of the phase coherence condition is
satisfied throughout this mass range.

The strategy outlined above assumes a macroscopic build-up, for which
$\alpha(t)$ and $\varphi(t)$ change significantly during the run.
For smaller resonance tunes (for exmaple of the order of, or below, the upper limits 
obtained in Ref.~\cite{Karanth2023}, both angles change only slightly within $\tau_a$, and
Eqs.~\eqref{eq:dalpha} and~\eqref{eq:dphi} reduce to constant slopes,
\begin{align}
  \frac{\dd\alpha_k}{\dd t} &\simeq 2\pi f_{\rm rev}\,
    (\epsilon_{\rm wind}-\epsilon_{\rm ac})\cos(\delta_k-\phi_a)\,,
  \label{eq:dalpha_lin}\\[4pt]
  \frac{\dd\varphi_k}{\dd t} &\simeq -2\pi f_{\rm rev}\tan\alpha_0\,
    (\epsilon_{\rm wind}+\epsilon_{\rm ac})\sin(\delta_k-\phi_a)\,,
  \label{eq:dphi_lin}
\end{align}
where $\delta_k$ is the known phase offset of bunch $k$, $\phi_a$ the common
unknown phase relative to the axion field, and $\alpha_0$ the initial
out-of-plane angle. Below the experimental sensitivity, both slopes are
consistent with zero.
Since $\dd\varphi/\dd t \propto \tan\alpha$, a first-order phase walk requires
$\alpha_0 > 0$. An initial tilt of the
polarization out of the ring plane is therefore mandatory.
A combined fit of Eqs.~\eqref{eq:dalpha_lin} and~\eqref{eq:dphi_lin} over all
bunches, with $\epsilon_{\rm wind}$, $\epsilon_{\rm ac}$ and $\phi_a$ as free
parameters, yields either values for both resonance tunes or, if both slopes are
consistent with zero, correlated limits in the
$(\epsilon_{\rm wind},\epsilon_{\rm ac})$ plane. These limits hold without the
assumption that one of the couplings vanishes, which is required when only the
build-up is measured. A quantitative estimate of the statistical uncertainty
depends on the polarimeter performance, the in-plane polarization lifetime and 
the number of bunches~(see Sect.~\ref{sect:statistics}).

\section{Disentanglement via the electric-to-magnetic field ratio}
\label{sec:EB}

A second, independent approach exploits the fact that $\epsilon_{\rm ac}$ and
$\epsilon_{\rm wind}$ depend differently on the ring fields and the beam momentum: $\epsilon_{\rm ac}
\propto (E - c\beta B)$ is proportional to the effective electric field in the
particle rest frame, while $\epsilon_{\rm wind} \propto \omega_a\beta$ depends
on the beam velocity.  In a combined electric-magnetic
storage ring (see, \emph{e.g.}, \cite{CPEDM2021}), $E$, $B$ and $\beta$ can in principle be varied while
maintaining the resonance condition at a given ring radius $R$.
This changes the relative contributions of $\epsilon_{\rm ac}$ and
$\epsilon_{\rm wind}$.  

\subsection{Constraints from ring radius and resonance condition}

The following discussion uses a circular storage ring with radius $R$ to illustrate the method.
In reality, ring geometries are more complex. Although this alters the complexity of the formulas, it does not affect the main strategy.

In a circular storage ring the relation between the electric and magnetic fields $E$ and $B$, the particle momentum $p$ and the ring radius $R$ is given by:
\begin{equation}
  E_{\rm eff} \equiv E - c\beta B = - \frac{p\,\beta c}{qR}\,.
  \label{eq:orbit}
\end{equation}
The negative sign originates from the definition $E_{\rm eff}>0$ pointing in direction of $\hat{e}_x$.

The resonance condition is given by Eq.~\eqref{eq:Oprec}:
\begin{equation}
  G B - \Bigl(G - \frac{1}{\gamma^2 - 1}\Bigr)\frac{\beta E}{c}
   = - \frac{\omega_a m}{q}\,.
  \label{eq:resonance}
\end{equation}

Eqs.~\eqref{eq:orbit} and~\eqref{eq:resonance} constitute two linear
equations in $E$ and $B$. The free parameter here is $p$ (together with the relativistic factors $\beta$ and $\gamma$). For each value of $p$, the two constraints fix a unique operating point
$(E, B, p)$ with the given precession frequency $\omega_a$.  The set of all such points forms a one-parameter family parameterized by $p$, tracing a curve in the $(E, B)$ plane.

Solving Eqs.~\eqref{eq:orbit} and~\eqref{eq:resonance} simultaneously for $E$ and
$B$ gives the explicit recipe for the required fields as a function of the chosen
beam momentum $p$:
\begin{align}
  B(p) &= - \frac{p}{qR\,(G+1)}
    \left[\frac{\gamma\,\omega_a R}{\beta c} + G\gamma^2\beta^2 - 1\right]\,,
  \label{eq:Bofp}\\[6pt]
  E(p) &= - \frac{p\,\gamma}{qR\,(G+1)}
    \left(\omega_a R + G\gamma\beta c\right)\,,
  \label{eq:Eofp}
\end{align}
where $\gamma$ and $\beta$ are functions of $p$ via $p = \gamma m\beta c$.  
For $E(p_0)=0$ and $\omega_a R = - G\gamma\beta c$ one gets the settings for a purely magnetic storage ring: $p_0 = - m R \omega_a / G$ and $B(p_0) = p_0 / (q R)$. 

\begin{table}
\caption{\label{tab:particles}%
  Particle properties used in this work.
  Masses and anomalous magnetic moments $G$ are from
  PDG~\cite{Workman:2022ynf} and CODATA~\cite{RevModPhys.93.025010}.}
\begin{ruledtabular}
\begin{tabular}{lcrrc}
  Particle  & Symbol          & $m$\;(MeV/$c^2$)   & $G$                   & $q/e$ \\
  \hline
  Proton    & $p$             & $938.272$      & $1.793$                & $+1$  \\
  Deuteron  & $d$             & $1875.613$      & $-0.1430$                & $+1$  \\
  Helium-3  & $^3\mathrm{He}$ & $2808.392$    & $-4.184$                & $+2$  \\
\end{tabular}
\end{ruledtabular}
\end{table}

\begin{figure*}
\includegraphics[width=\textwidth]{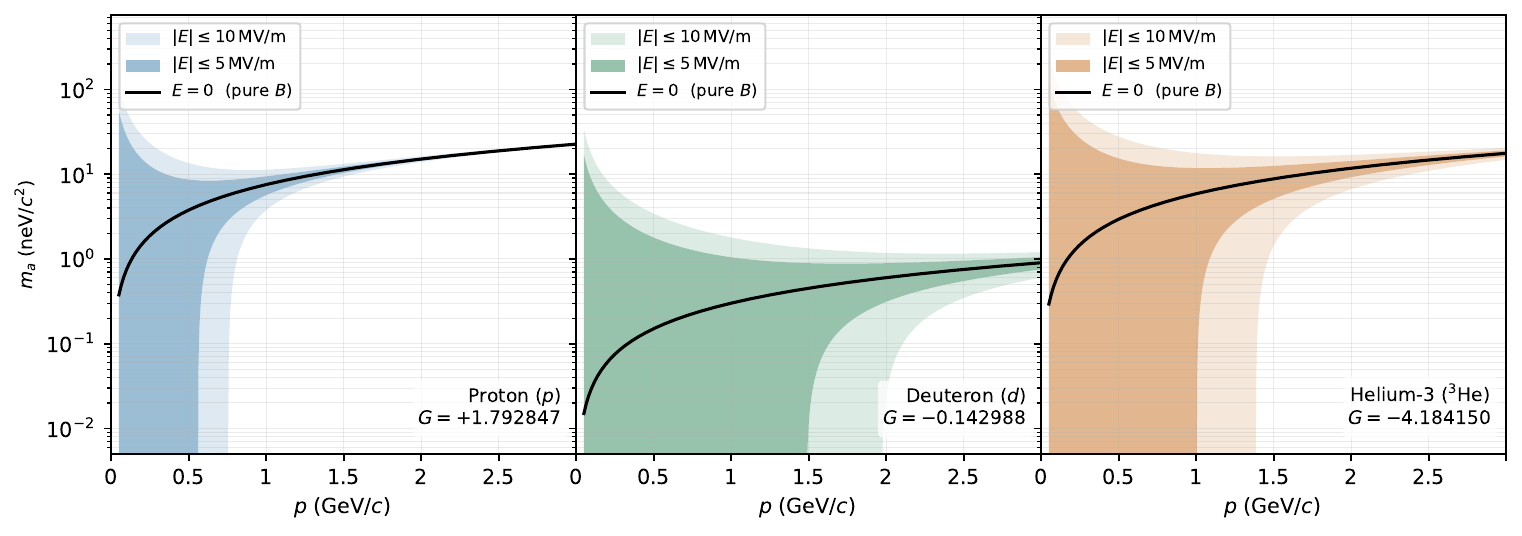}
\caption{Accessible range of axion masses for various particles and momenta up to $p = \SI{3}{GeV/c}$ assuming a ring radius of $R=\SI{50}{m}$. The solid black line corresponds to a purely magnetic ring. The shaded areas show the additional range accessible in a combined ring with electric fields limited to $|E| < \SI{5}{MV/m}$ and $|E| < \SI{10}{MV/m}$.}
\label{fig:ma_range}
\end{figure*}

\begin{figure*}
\includegraphics[width=\textwidth]{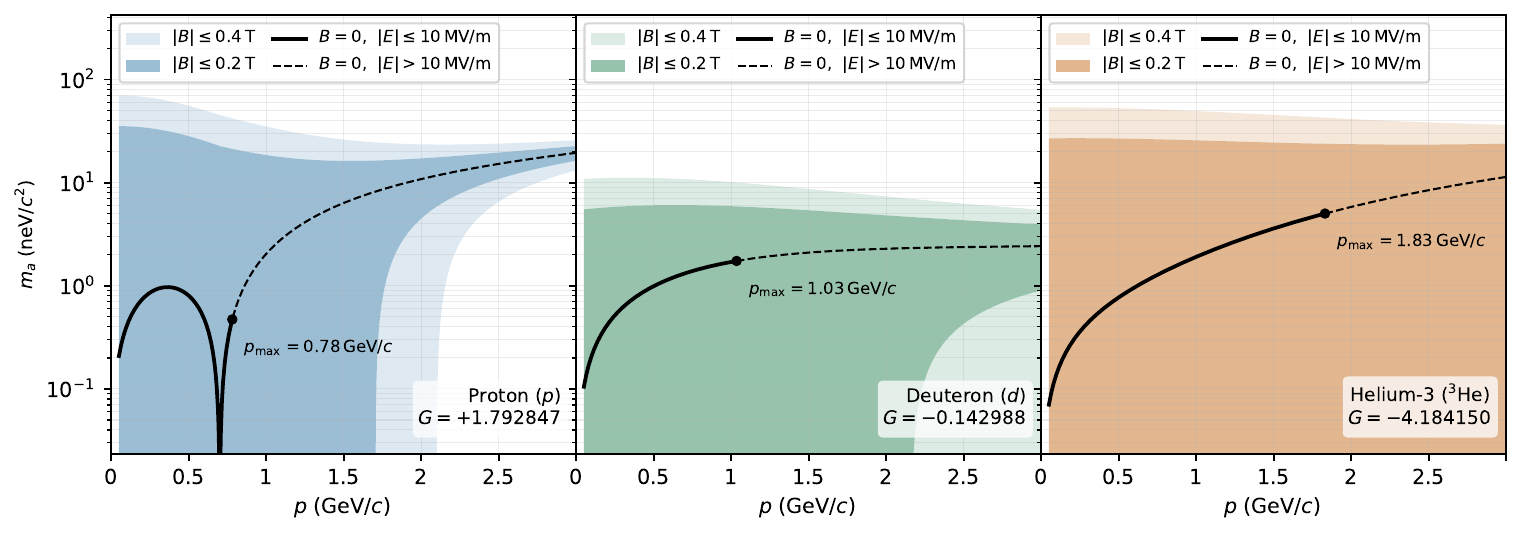}
\caption{Similar to Fig.~\ref{fig:ma_range}, but starting off a purely electrostatic ring. The black line shows the resonance condition for $B = 0$, with the corresponding electric field $E = p\beta c/(qR)$. The line is drawn solid up to the momentum $p_{\max}$ (dot) at which $E$ reaches $10\,\mathrm{MV/m}$, and dashed where larger fields would be required. For protons the resonant mass vanishes at the frozen-spin momentum $p = mc/\sqrt{G} \simeq 0.70\,\mathrm{GeV}/c$. The shaded bands
indicate the mass range accessible at fixed momentum by adding a vertical
magnetic field of $|B| \leq 0.2\,\mathrm{T}$ (dark) or $|B| \leq
0.4\,\mathrm{T}$ (light).
}
\label{fig:ma_range_B0}
\end{figure*}

\subsection{Resonance tunes}

$\epsilon_{\rm ac}$ and $\epsilon_{\rm wind}$ need to be expressed as functions of
$p$, $R$ and $\omega_a$. For a circular orbit with radius $R$, the revolution frequency is
\begin{equation}
  f_{\rm rev} = \frac{\beta c}{2\pi R}\,.
  \label{eq:frev}
\end{equation}
Together with Eq.~\eqref{eq:orbit} one can rewrite the Eqs.~\eqref{eq:eac}
and~\eqref{eq:ewind}:
\begin{align}
  \epsilon_{\rm ac}
  &= \frac{d_{\rm ac}}{S\hbar} \cdot \frac{E_{\rm eff}}{4\pi f_{\rm rev}}
   = - \frac{d_{\rm ac}}{S\hbar} \cdot \frac{p\,\beta c/(qR)}{2\beta c/R}
  \nonumber\\
  &= - \frac{d_{\rm ac}}{2S\hbar q}\,p\,,
  \label{eq:eac_p}\\[6pt]
  \epsilon_{\rm wind}
  &= - \frac{C_N\,a_0\,\omega_a\,\beta}{2f_a S} \cdot \frac{1}{4\pi f_{\rm rev}}
   = - \frac{C_N\,a_0\,\omega_a\,\beta}{2f_a S} \cdot \frac{R}{2\beta c}
  \nonumber\\
  &= - \frac{C_N\,a_0\,R}{4f_a S c} \,\omega_a\,.
  \label{eq:ewind_R}
\end{align}
In $\epsilon_{\rm ac}$ the factors $\beta$ and $R$ cancel, leaving a
result proportional to $p$. In $\epsilon_{\rm wind}$ the $\beta$ from
$\vec\Omega_{\rm wind}$ cancels against the $\beta$ in $f_{\rm rev}$, leaving a
quantity that depends only on $\omega_a$ and $R$ and is therefore constant for fixed $\omega_a$ and $R$.

The full resonance tune $\epsilon_{\rm wind} - \epsilon_{\rm ac}$ that drives the build-up $d\alpha/dt$ then varies as
\begin{equation}
  \mathcal{A}(p)
  = \epsilon_{\rm wind} + \frac{d_{\rm ac}}{2S\hbar q}\,p\,,
  \label{eq:Atotal}
\end{equation}
which is linear in $p$ with a proportionality coefficient that defines $d_\mathrm{ac}$.
Therefore, measuring $\mathcal{A}$ at two or more values of $p$ provides a linear system from which both parameters,
$\epsilon_{\rm wind}$ and $d_{\rm ac}$, can be extracted independently.

\subsection{Accessible axion mass range}
\label{sec:ma_range}

For the four particle species listed in Table~\ref{tab:particles} Fig.~\ref{fig:ma_range} shows the axion mass window reachable in a combined electric-magnetic storage ring with a radius of $R = \SI{50}{m}$ as a function of beam momentum. 

\paragraph{Pure magnetic field.}
Setting $E = 0$ in Eq.~\eqref{eq:Eofp} gives the resonant axion mass for a purely magnetic ring:
\begin{equation}
  (m_ac^2)^{E=0} = \frac{\hbar\,|G|\,\gamma\beta c}{R} = \frac{\hbar\,|G|\,p}{m R}\,.
  \label{eq:ma_purB}
\end{equation}
This is the solid black line in each panel of Fig.~\ref{fig:ma_range}.  It grows almost linearly with~$p$ for $p \gg mc$ and scales with $R^{-1}$.

\paragraph{Electric-field range.}
An additional, non-zero electric field extends the accessible axion mass.
Differentiating Eq.~\eqref{eq:Eofp} yields
\begin{equation}
  \frac{\partial E}{\partial \omega_a} = -\frac{p\,\gamma}{q\,(G+1)}\,,
  \label{eq:dEdwa}
\end{equation}
so the accessible half mass range around the pure-$B$ resonance at field limit $E_{\rm lim}$ is
\begin{equation}
  \Delta(m_a c^2) = \frac{\hbar\,|q\,(G+1)|}{p\,\gamma}\,\Delta E,
  \label{eq:Delta_ma}
\end{equation}
shown as the shaded bands in Fig.~\ref{fig:ma_range}.
The relative mass range $\Delta m_a / m_a^{E=0}$
decreases as $\gamma^{-2}$ at high momentum.
For example, for protons with $R = \SI{50}{m}$ and $E_{\rm lim} = \SI{5}{MV/m}$, the full accessible band spans $\approx\SI{3.8}{neV}/c^2$ (50\% of $m_a^{E=0}$) at $p = \SI{1}{GeV/c}$, and shrinks to $\approx\SI{0.55}{neV}/c^2$ ($\approx 2\%$) at $p = \SI{3}{GeV/c}$.

\paragraph{Phase coherence.}
The phase-coherence condition of Sec.~\ref{sec:disentangle} --- that the observation time remains short compared to $\tau_a$ --- also applies here, but its consequence is qualitatively different.
There, the run had to stay phase-coherent while the time dependence of $\alpha(t)$ and $\varphi(t)$ was measured, resulting in an absolute mass cutoff for a chosen minimum run length.
Here, the total build-up is measured ($d\alpha/dt \propto \A(p)$, Eq.~\eqref{eq:dalpha}), and the figure of merit is $\A(p)\tau_a$: the larger the coupling is, the smaller $\tau_a$ can be.

\paragraph{Magnetic field.}
At the pure-$B$ operating point Eq.~\eqref{eq:Bofp} reduces to the standard magnetic rigidity $B = p/(qR)$, so the maximum reachable momentum scales as $p_{\rm max} = qB_{\rm max}R$.
The momentum range shown in Fig.~\ref{fig:ma_range} requires $B \lesssim \SI{0.07}{T}$ at $p = \SI{1}{GeV/c}$ and $B \lesssim \SI{0.2}{T}$ at $p = \SI{3}{GeV/c}$, which is comfortably within standard magnet technology.
Higher fields, up to superconducting magnet technology, would in principle permit correspondingly higher momenta and thus higher axion masses. However, such masses lie beyond the reach set by the phase-coherence condition.

For completeness, Fig.~\ref{fig:ma_range_B0} shows the accessible mass range starting from a purely electrostatic ring ($B = 0$) and adding a vertical magnetic field. 
The electrostatic baseline follows from Eq.~\eqref{eq:resonance} with $B = 0$ and $E = p\beta c/(qR)$:
\begin{equation}
  (m_ac^2)^{B=0} = |G\,\beta^2\gamma^2 - 1|\,\frac{\hbar\,\beta c}{\gamma R}\,.
  \label{eq:ma_pureE}
\end{equation}
The shaded bands result from differentiating Eq.~\eqref{eq:resonance} with respect to $B$ at fixed $p$ and $R$, using Eq.~\eqref{eq:orbit} to account for the change in $E$:
\begin{equation}
  \Delta(m_a c^2) = \frac{\hbar\,|q(G+1)|}{\gamma^2 m}\,\Delta B\,.
  \label{eq:Delta_ma_B}
\end{equation}

\subsection{Achievable separation precision}
\label{sec:separation_precision}

The E-field half-bandwidth $\Delta m_a$ identified in Sec.~\ref{sec:ma_range} maps one-to-one onto a momentum lever arm $\Delta p$ available at fixed $m_a$.
This controls how well $\epsilon_{\rm wind}$ and $d_{\rm ac}$ can be disentangled once $\mathcal{A}(p)$, Eq.~\eqref{eq:Atotal}, is measured at two momenta $p_1$ and $p_2\equiv p_1+\Delta p$ with independent amplitude uncertainty $\sigma_A$. Writing $\mathcal{A}_i = \epsilon_{\rm wind}-k\,p_i$ with $k = d_{\rm ac}/(2S\hbar q)$ and solving the resulting linear system for $k$ and $\epsilon_{\rm wind}$, standard error propagation gives
\begin{align}
  \sigma_{k} &= \frac{\sqrt{2}\,\sigma_A}{\Delta p}\,,
  \label{eq:sigma_k}\\[4pt]
  \sigma_{\epsilon_{\rm wind}} &= \sigma_A\,\frac{\sqrt{p_1^2+p_2^2}}{\Delta p}
  \;\xrightarrow{\;\Delta p \ll p\;}\;
  \sqrt{2}\,\frac{p}{\Delta p}\,\sigma_A\,,
  \label{eq:sigma_ewind}
\end{align}
with $p = (p_1+p_2)/2$. The extracted $\epsilon_{\rm wind}$ uncertainty scales with the combination $\sigma_A\,p/\Delta p$, its separation quality is set by the fractional lever arm $\Delta p/p$. $\sigma_k$ scales as $\sigma_A/\Delta p$. The two-point measurement at the extremes of the accessible range is the best case for both quantities: splitting the same total measurement time and momentum range across more points only degrades $\sigma_k$ and $\sigma_{\epsilon_{\rm wind}}$.

Table~\ref{tab:leverarm} lists the momentum lever arm $\Delta p$ obtained by solving the resonance condition, Eq.~\eqref{eq:Eofp}, for the proton example of Sec.~\ref{sec:ma_range} ($R=\SI{50}{m}$, $E_{\rm lim}=\SI{5}{MV/m}$), together with the resulting degradation factor $\sqrt2\,p/\Delta p$. At $p=\SI{1}{GeV/c}$ the accessible range is one-sided, extending only up to $p\approx\SI{1.19}{GeV/c}$, with no corresponding bound on the low side within the $E$-field limit.

\begin{table}
\caption{\label{tab:leverarm}%
  Momentum lever arm $\Delta p$ and degradation factor $\sqrt2\,p/\Delta p$ (Eqs.~\eqref{eq:sigma_k}, \eqref{eq:sigma_ewind}) for protons, $R=\SI{50}{m}$, $E_{\rm lim}=\SI{5}{MV/m}$.}
\begin{ruledtabular}
\begin{tabular}{ccc}
  $p$ & $\Delta p$ & $\sqrt2\,p/\Delta p$ \\
  \hline
  \SI{1}{GeV/c} & \SI{0.19}{GeV/c} (one-sided) & 7.5 \\
  \SI{3}{GeV/c} & \SI{0.07}{GeV/c} & 60 \\
\end{tabular}
\end{ruledtabular}
\end{table}

This assumes uncorrelated, statistics-limited amplitude uncertainties at the two momentum points, systematic effects are not included. A dedicated sensitivity study, folding in a realistic $\sigma_A$ and systematic uncertainties, needs to be performed for a specific ring design. 

\section{Practical considerations}
\label{sec:practical}

\subsection{Spin resonances}
Each point in Figs.~\ref{fig:ma_range} and~\ref{fig:ma_range_B0} corresponds to a separate fixed-momentum operating point. For given $m_a$ and $R$, the chosen momentum determines both the field configuration, Eqs.~\eqref{eq:Bofp} and~\eqref{eq:Eofp}, and the spin tune $\nu_s = \omega_a R/(\beta c)$. Operating points at or close to a spin resonance, $\nu_s = k$ or $\nu_s = k\pm Q_y$, have to be avoided. An integer resonance $\nu_s = k$ occurs for a given $m_a$ only at the momentum with $\beta = m_a c R/(k\hbar)$, so that the same mass can be probed at a different momentum within the accessible band. Intrinsic resonances can in addition be shifted by the choice of the betatron tunes.

\subsection{Momentum-dependent systematics}
The separation in Eq.~\eqref{eq:Atotal} relies on the $p$-dependence of $\A$, so any instrumental effect that changes with momentum enters the slope directly. The most direct one is the polarimeter calibration: the build-up is calculated from an asymmetry proportional to the analyzing power $A(p)$, and a momentum-dependent calibration error produces a slope that mimics a contribution of $d_{\rm ac}$. The polarimeter figure of merit also varies with $p$, so that the uncertainties $\sigma_\A$ at the two momenta generally differ from the equal-error assumption of Sec.~\ref{sec:separation_precision}. 
A further effect is the tilt of the invariant spin axis by imperfection fields, which changes the projection of both torques onto the build-up direction and varies with $p$. Finally, the sextupole settings that maximize the in-plane
polarization lifetime have to be re-optimized at each momentum. The build-up therefore has to be normalized to the in-plane polarization measured in the same run.

\subsection{Statistical accuracy considerations}
\label{sect:statistics}
Starting form equations~\ref{eq:dalpha_lin} and \ref{eq:dphi_lin} the 
statistical accuracy can be evaluated using the procedure discussed in
~\cite{Pretz2020} for the vertical build-up $\dot{\alpha}_k$ and \cite{Eversmann2016} for the spin tune measurement $\dot{\varphi}_k$.
	\begin{eqnarray*}
	\sigma_{\dot{\alpha_k}} &=& \sqrt{\frac{24}{Nf}} \, \frac{1}{PAT} \, \frac{1}{\cos(\alpha_0)} \, \frac{1}{\sqrt{2}} \, , \\
		\sigma_{\dot{\varphi_k}} &=& \sqrt{\frac{48}{Nf}} \, \frac{1}{P A T} \,  \frac{1}{\cos(\alpha_0)}\, \frac{1}{\sqrt{2}} \, .
	\end{eqnarray*}
The additional factor $\cos(\alpha_0)$ takes into account, that the initial direction of the polaristation vector is out of the horizontal plane. The factor $1/\sqrt{2}$ takes into account that four bunches have to be used because of the unknown phase of the axion field.
	
 Assuming that the polarisation is determined via proton carbon scattering with an analysing power $A$ and using the numerical values listed in table~\ref{tab:t1}
 allows one to evaluate
	the errors on $\epsilon_{\rm wind}$ and $d_{\rm ac}$.
	\begin{table}[h!]
 \begin{ruledtabular}
		\begin{tabular}{lll}
			Variable & Meaning & typical value \\
			\hline
			$A$        & analyzing power &  0.6 \\
			$P$        & beam polarization & 0.8 \\
			$T=$ &      &                       \\
           $\mbox{min}(T_{\rm sct},\tau_c)$  & spin coh. time/coh. length & \SI{10}{s} \\
			$N $       & number of particles in cycle & $4 \cdot 10^{10}$ \\
			$f $        & polarimeter efficiency  &  0.005 \\
			$p$         & momentum  &  $[0.3,1]$ \si{GeV/c} \\
		\end{tabular}
		\caption{Variables with typical values. \label{tab:t1}}
 \end{ruledtabular}
	\end{table}
 
 For $\alpha_0 = \pi/4$ the highest accuracy is reached. For one accelerator cycle one finds:
 \begin{eqnarray*}
	\sigma_{\epsilon_{\rm wind}} &=& \SI{8.3e-12}{} \, ,\\
	\sigma_{d_{\rm ac}} &=& \SI{1.2e-25}{{\it e} \, cm} \, . 
 \end{eqnarray*}

For the second method 
 using equation~\ref{eq:dalpha} alone at two momenta $p_1$ and $p_2$ one finds
 for the  uncertainties
 \begin{eqnarray}
    	\sigma_{\epsilon_{\rm wind}} &=&  \sqrt{p_2^2 \sigma_{\mathcal{A}_1}^2 + p_1^2 \sigma_{\mathcal{A}_2}^2} \, \frac{1}{p_2 - p_1} \, ,\\	
    	\sigma_{d_{\rm ac}} &=&  \sqrt{\sigma_{\mathcal{A}_1}^2 +\sigma_{\mathcal{A}_2}^2} \, \frac{1}{p_2 - p_1} \, 2 S \hbar q  
\end{eqnarray}
     with $\mathcal{A}_{1,2} \equiv \mathcal{A}(p_{1,2})$.
     The numerical values are
     	\begin{eqnarray*}
		\sigma_{\epsilon_{\rm wind}} &=& \SI{4.0e-11}{} \, ,\\
		\sigma_{d_{\rm ac}} &=& \SI{8.5e-25}{{\it e} \, cm} \, .
	\end{eqnarray*}
	This corresponds to a running time of two cycles with
	momentum 0.3 and 1 \si{GeV/c}, respectively.

\subsection{Resonance stabilization}
Both strategies require the spin precession to stay on resonance with the
probed axion frequency over the whole run. At COSY, the spin precession has been
phase-locked to an external rf reference by a feedback loop that determines
the in-plane precession phase in real time and corrects the revolution
frequency of the beam, and with it the spin tune~\cite{JEDI:2017bnp}.
The feedback is driven by a single reference bunch.
For the method of Sec.~\ref{sec:EB}, where only the build-up is measured and
the polarization starts in the ring plane, any bunch can serve as reference.
For the phase-walk method, the feedback loop cancels the axion-induced phase
walk in the reference bunch together with any spin-tune drift. The correction
of the revolution frequency acts on all bunches alike and induces a common
phase correction. The observable is then the phase walk relative to the
reference bunch.
Alternatively, the feedback can be used to set the initial phase and be
switched off during the measurement~\cite{JEDI:2018txs}, provided the
spin-tune drift over the run duration is small.

\section{Summary and outlook}
\label{sec:summary}

Storage rings are intrinsically more sensitive to $\epsilon_{\rm wind}$, which is enhanced by a factor $\beta/\beta_\odot \sim 10^3$ relative to laboratory-frame experiments.
The two complementary strategies discussed in this paper allow the oscillating-EDM and axion-wind couplings to be separated. 

First, the two couplings enter the vertical polarization build-up and the spin-precession phase walk as difference and sum, respectively (Eqs.~\eqref{eq:dalpha} and~\eqref{eq:dphi}), even though they are degenerate in the build-up alone. Running at resonance at fixed energy and tracking both the build-up and the phase-walk amplitude simultaneously allows individual extraction of $\epsilon_{\rm wind}$ and $\epsilon_{\rm ac}$ up to 
an overall sign ambiguity.
The method requires the run duration to be short compared to the axion coherence time $\tau_a$ and the in-plane polarization lifetime to be sufficiently long.

Second, $\epsilon_{\rm ac}$ is proportional to the beam momentum $p$ while
$\epsilon_{\rm wind}$ is independent of $p$ at a given ring and a fixed resonance frequency (Eqs.~\eqref{eq:eac_p} and~\eqref{eq:ewind_R}). The
total resonance amplitude is then linear in $p$ (Eq.~\eqref{eq:Atotal}), enabling a clean separation.

Neither approach is applicable with the scan technique employed in
Ref.~\cite{Karanth2023}.  
Both are however feasible at future combined electric-magnetic storage rings planned for permanent-EDM searches~\cite{CPEDM2021}, which would extend
the accessible axion mass range to the full window shown in current exclusion
plots~\cite{Karanth2023}.  
Using different beam species would further allow to study the dependence on the anomalous magnetic moment $G$ and on nuclear structure, providing additional cross-checks on the disentanglement.

It should be noted, that a separate theoretical question concerns whether the fermion coupling truly generates a physical oscillating EDM. 
In Ref.~\cite{DiLuzio2024} a previously overlooked axion boundary term has been identified, required to restore the axion shift symmetry, that suppresses the fermion-coupling EDM in the slow-oscillation regime and produces an exact cancellation in the static limit — a conclusion reached independently in Ref.~\cite{Smith2024} via a generalised Schiff theorem. However, in \cite{Smith2024} it is further argued that for nucleons a residual axionic EDM can survive the screening, potentially at a level exceeding the gluon-coupling contribution. The implications for storage-ring searches therefore remain an open question: if the fermion-coupling EDM is suppressed, the oscillating EDM signal would map more cleanly onto the gluon coupling alone, simplifying the disentanglement problem; if instead a nucleon-level axionic EDM persists, the two contributions would remain entangled.
Since the methods presented here determine $\epsilon_{\rm ac}$ independently of $\epsilon_{\rm wind}$, they can provide direct experimental input to this question.

\begin{acknowledgments}
We thank our colleagues from the JEDI Collaboration for many valuable and helpful discussions.
\end{acknowledgments}


\bibliography{literature}

\end{document}